\documentclass[9pt,twocolumn,twoside]{optica}
\setboolean{shortarticle}{false}
\setboolean{minireview}{false}
\usepackage{adjustbox}
\usepackage{graphicx}
\usepackage{comment}
\usepackage{amsmath}
\usepackage{lipsum}
\usepackage{float}
\usepackage{nicefrac}
\usepackage{soul}
\usepackage{cancel}
\usepackage[normalem]{ulem}
\usepackage{multirow,tabularx}
\usepackage{dblfloatfix}
\usepackage{nidanfloat}

\title{Mode Selective Image Upconversion over Turbulence}

\author[1,2]{He Zhang}
\author[1,2]{Santosh Kumar}
\author[1,2,*]{Yu-Ping Huang}

\affil[1]{Department of Physics, Stevens Institute of Technology, Hoboken, NJ, 07030, USA}
\affil[2]{Center for Quantum Science and Engineering, Stevens Institute of Technology, Hoboken, NJ, 07030, USA}

\affil[*]{Corresponding author: yuping.huang@stevens.edu}

 \dates{Compiled \today}

\begin{abstract}
We experimentally study a nonlinear optical approach to selective manipulation and detection of structured images mixed with turbulent noise. Unlike any existing adaptive-optics method by applying compensating modulation directly on the images, here we account for the turbulence indirectly, by modulating only the pump driving the nonlinear process but not the images themselves. This indirect approach eliminates any signal modulation loss or noise, while allowing more flexible and capable operations. Using specifically sum frequency generation in a lithium niobate crystal, we demonstrate selective upconversion of Laguerre-Gaussian spatial modes mixed with turbulent noise. The extinction reaches $\sim$40 dB without turbulence, and maintains $\sim$20 dB in the presence of strong turbulence. This technique could find utilities in classical and quantum communications, compressive imaging, pattern recognition, and so on. 

\end{abstract}

\setboolean{displaycopyright}{true}

\begin{document}

\maketitle

%%%%%%%%%%%%%%%%%%%%%%%%%%  body  %%%%%%%%%%%%%%%%%%%%%%%%%%
\section{Introduction}
Nonlinear optics (NLO) underpins optical parametric oscillation \cite{Miller65,Khoury18}, parametric downconversion \cite{Guo16,YMS17}, harmonic generation \cite{Delaubert07,Pereira17}, sum-frequency generation \cite{Vasilyev14,Maestre18}, four wave mixing \cite{Boyer08,Ding12}, etc., with wide applications in optical communications \cite{Boyd03,Lassen07}, biomedical engineering \cite{Cheng11}, metrology \cite{Glasser08}, and quantum information \cite{Hallett:18}. For optical signal processing and detection, NLO techniques can offer significant advantages over their linear-optics counterparts \cite{Delaubert07,Lanning17,Demur18}, as demonstrated repeatedly in temporal mode-selective frequency conversion  \cite{Benjamin11,Eckstein11,Reddy13,YuPing13,Brecht13,Allgaier17,QPMS2017}, lossless photon shaping \cite{Koprulu11}, spiral phase contrast imaging of the edges \cite{Xiaodong18}, and field-of-view enhancement \cite{Maestre18,liu_2019}. To capitalize on the rich spatial features of light, frequency upconversion has been utilized for mode-selective detection of spatially orthogonal signals in few-mode waveguides \cite{Vasilyev14,Vasilyev17}, and more recently in nonlinear crystals to selectively convert overlapping Laguerre-Gaussian (LG) and Hermite-Gaussian (HG) modes \cite{Santosh19,sephton2019}.

In this paper, we further those studies towards field applications and demonstrate mode-selective upconversion and detection of overlapping images mixed with strong turbulence noise. Spatially structured signals, such as those in LG and HG modes, are useful for quantum information processing, remote sensing, and so on \cite{Lassen07,Brecht13,Pang18,fontaine_2019}. Yet they are quite susceptible to multiscattering and turbulence \cite{Malik:12,Shuhui18,Zhou:19,rubinsztein-dunlop_roadmap_2017}. The ability to distinguish and separately detect or manipulate those disturbed modes is essential to recover the information they carry, as critical for pattern recognition \cite{Kolmogorov,Ren:14}, free-space communications \cite{Sit:17,Bozinovic13,Vasilyev14,Huang15,Ren:14,Vasilyev17}, compressive sensing \cite{Howland13,Xiaojun11}, and LiDAR \cite{Amin18}. To this end, a variety of phase compensation methods \cite{Qassim14, Mair01} have been demonstrated using adaptive optics \cite{krenn_communication_2014,rodenburg_simulating_2014,gemayel_cross-talk_2016,li_adaptive_2017}, via feedback \cite{Shuhui18}, feed-forward, or machine learning \cite{Glasser18,Fan19}. All of them rely on applying phase or amplitude modulations directly onto the signals, which could induce excessive noise and noise while subject to limited capabilities.   

Here, we explore a rather distinct, indirect approach via turbulence-compensating upconversion. Instead of modulating the images, we account for the turbulence using mode-modulated pump to drive the image upconversion, which can select overlapping images despite strong turbulence. This indirect compensation approach avoids any modulation loss or noise added to the signals while allowing more flexible operations.

As a concrete example, here we utilize sum frequency (SF) generation in a lithium niobate crystal, where the signal and pump interact while experiencing strong diffraction. By preparing the pump in an optimized mode, overlapping signals can be upconverted selectively according to their orthogonal spatial modes. As any turbulence can be described by a unitary linear-optical transformation, distinct modes passing through turbulence will have distorted profiles but  will nonetheless remain orthogonal. This allows us to design the pump via adaptive feedback control, to selectively upconvert certain modes over others, even when their intensity profiles are strongly distorted by turbulence to totally unrecognizable. We demonstrate the selective upconversion of one LG modes against another with up to 40 dB extinction without turbulence, and up to 21 dB under strong turbulence. Our experimental results match fairly well with numerical simulations without using any fitting parameter. 

\section{Theory}

The LG modes can be written in the cylindrical coordinate system as \cite{lanning_gaussian_2017}
\begin{multline}%\psi \equiv 
LG_l^{p}(r,\phi,z)=\frac{C_{lp}}{w(z)}\left(\frac{r\sqrt{2}}{w(z)}\right)^{|l|}\exp\left(\frac{-r^{2}}{w^{2}(z)}\right) \exp(-ikz)\\ \times L_{p}^{|l|}\left(\frac{2r^{2}}{w^{2}(z)}\right) \exp\left(-ik\frac{r^{2}}{2R(z)}\right)\exp(-il\phi)\exp(i\zeta(z)),
\end{multline}
where $r=\sqrt{x^2+y^2}$ is the radial coordinate, $\phi=\arctan(y/x)$ is the azimuthal coordinate, ${\displaystyle C_{lp}}=\sqrt{\frac{2p!}{\pi(p+|l|)!}}$ is a normalization constant, ${\displaystyle w(z)}$ =$w_0$ $\sqrt{1 + (z/z_R)^2}$ is the beam radius at $z$, $w_0$ is the beam waist, $z_R$ = $\pi w_0^2/\lambda$ is the Rayleigh range, ${\displaystyle R(z)}$ = $z(1 + (z_R/z)^2)$ is the curvature radius of the beam, $\{\displaystyle L_{p}^{\vert l \vert}\}$ are the generalized Laguerre polynomials with the azimuthal mode index $l$ and the radial index $p$, $k$ = $2\pi n/\lambda$ is the wave number, and $\zeta(z)\equiv (2p+|l|+1) \arctan(z/z_R)$ is the Gouy phase at $z$.

\begin{figure*}[b]%!b
\centering %\linewidth
\includegraphics[width=15cm]{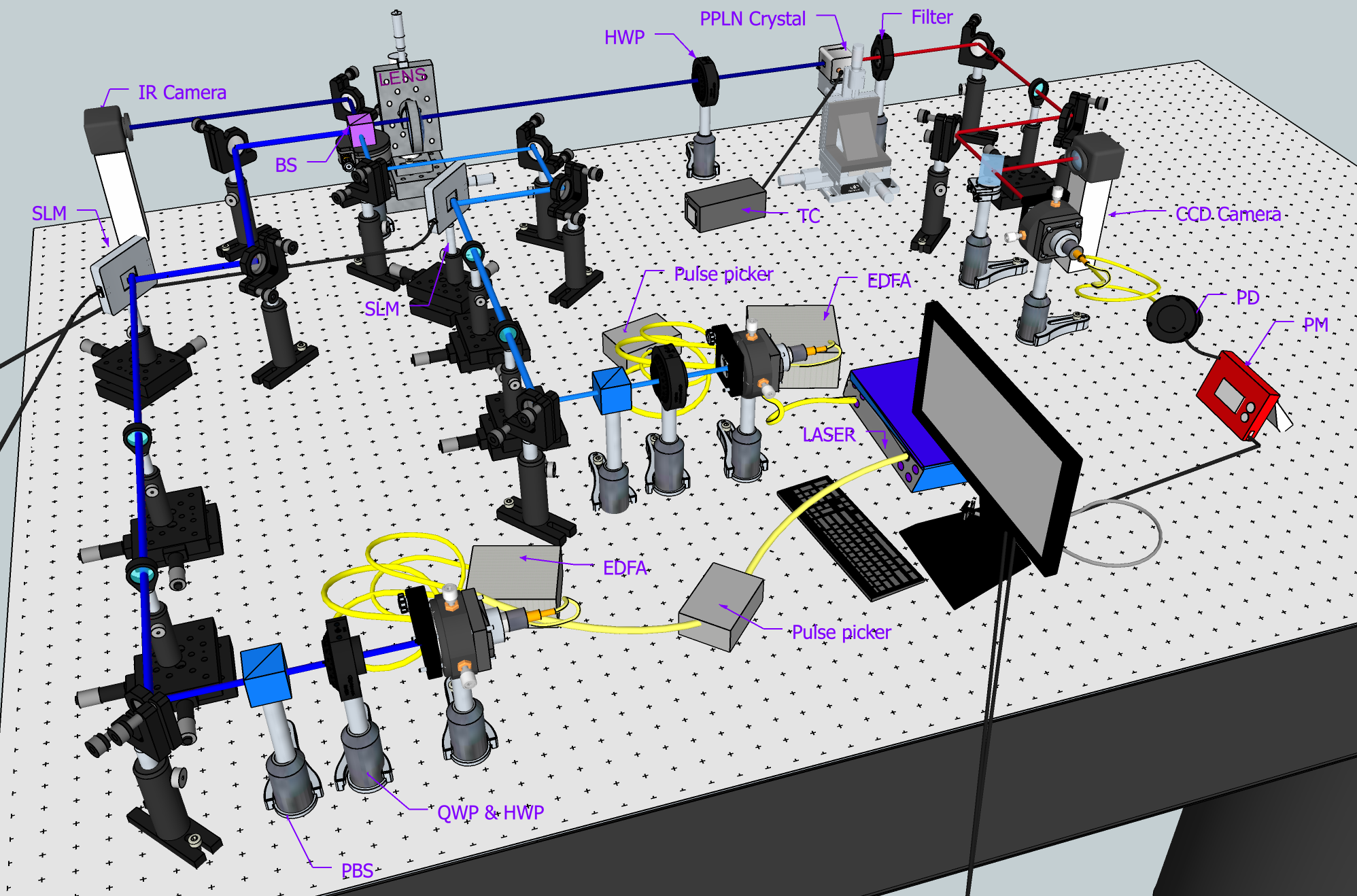}
\caption{Experimental setup. Two synchronized light pulse trains, each at 1544.9 and 1558.5 nm, are created through commonly referenced pulse pickers. The SLMs are used to create the desired spatial modes. The pulse trains are combined using a BS and passed through a temperature stabilized PPLN crystal. The generated SF light at 775.85 nm is filtered and coupled into a single mode fiber for detection using a power meter. Its result is fed via MATLAB to adaptively optimize the phase mask on pump SLM. EDFA: Erbium-doped fiber amplifier, QWP: Quarter waveplate, HWP: Half waveplate, BS: Beamsplitter, SLM: Spatial Light Modulator, PPLN crystal: Magnesium-doped periodic poled lithium niobate crystal, PD: Photodiode, PM: Powermeter, TC: Temperature Controller. 
} \label{ExpSetUp}
\end{figure*}

The SF is described under the slowly-varying-envelope approximation as: 
\begin{equation}
2i\kappa_{s}\partial_z\psi_{s}+(\partial_{x}^{2}+\partial_{y}^{2})\psi_{s}=-2\frac{\omega_{s}^{2}}{c^{2}}\chi^{(2)}\psi_{p}^{*}\psi_{f}e^{i\triangle\kappa z},
\label{eq2}
\end{equation}
%2
\begin{equation}
2i\kappa_{p}\partial_z\psi_{p}+(\partial_{x}^{2}+\partial_{y}^{2})\psi_{p}=-2\frac{\omega_{p}^{2}}{c^{2}}\chi^{(2)}\psi_{s}^{*}\psi_{f}e^{i\triangle\kappa z},\label{eq3}
\end{equation}
%3
\begin{equation}
2i\kappa_{f}\partial_z\psi_{f}+(\partial_{x}^{2}+\partial_{y}^{2})\psi_{f}=-2\frac{\omega_{f}^{2}}{c^{2}}\chi^{(2)}\psi_{p}\psi_{s}e^{-i\triangle\kappa z},
\label{eq4}
\end{equation}
where $\Delta\kappa=\kappa_{s}+\kappa_{p}-\kappa_{f}-2\pi/\Lambda$ is the momentum mismatching. $\kappa_{s}=\frac{n_{s}\omega_{s}}{c}$ , $\kappa_{p}=\frac{n_{p}\omega_{p}}{c}$ and $\kappa_{f}=\frac{n_{f}\omega_{f}}{c}$
 are the wave numbers of signal, pump and SF light, respectively. $\chi^{(2)}$ is the second-order nonlinear susceptibility and $\Lambda$ is the poling period of the nonlinear crystal. %The subscripts j=s,p,f denote the signal, pump, and SF light, respectively. 
$\psi_{s},\psi_p$, and $\psi_f$ are the electric fields for the signal, pump, and SF, respectively. The energy in the frequency upconversion process is conserved i.e., $\omega_p + \omega_s = \omega_f$. We use the standard split-step Fourier and adaptive step size methods to numerically solve Eqs.(\ref{eq2})-(\ref{eq4}) \cite{AGRAWAL2013}.

The total SF power is $P_f=2\epsilon_0 c n_f \int_0^{\infty} \int_0^{\infty}|\psi_f(x,y,z)|^2 dx dy,$ where $\epsilon_0$ is the permittivity and c is the speed of light in vacuum. In our experiment, however, the SF light is first coupled into a single mode fiber before detection. This is necessary to achieve both high selectivity and high collection efficiency, as critical for photon-starving and/or cascaded quantum applications \cite{Amin18,hughes2002}. The conversion extinction of two modes is then defined as the ratio of their converted SF power coupled into the fiber by the same pump. 

To study the turbulence effects, we use the Kolmogorov model to simulate the refractive index variation in the atmospheric turbulence \cite{Kolmogorov}. The power spectral density for the refractive-index fluctuations can be defined as 

\begin{equation}
\Phi_n(z,k)=0.0033 c_n^2(z) k^{-11/3}, 
\end{equation} 

with $1/L_0 << k <<1/l_0$. Here $k$ is the scalar wave number, $L_0$ and $l_0$ are the outer and inner scales for the turbulence, respectively and $c_n^2$ is the structure constant of the atmosphere at the propagation distance $z$. The strength of the turbulence is then approximately defined by Strehl ratio ($SR$) as 

\begin{equation}
SR = \frac{1}{1+(D/r_0)^{5/3}},
\label{SR}
\end{equation} 

where D is the diameter of the optical beam, $r_0$ is the Fried's parameter, with $r_0= 0.18 (\lambda^2/c_n^2 z)^{3/5}$ \cite{Kolmogorov,Shuhui18}. $SR$ varies between 0 and 1, with $SR$ = 0 indicating the maximum turbulence and $SR$=1 no turbulence. 

\section{Experimental Set Up}

The experimental setup is shown in Fig. \ref{ExpSetUp}. We create the pump and signal pulses by electro-optically modulating the outputs of continuous lasers, each at 1544.9 nm and 1558.5 nm. Those pulses are synchronized to a common reference radio-frequency source. Each pulse has a 200-ps full width at half maximum (FWHM) and 10-MHz repetition rate. The pulses are amplified using two separate Erbium-Doped Fiber Amplifiers (EDFAs). The pump's average power is 40 mW and peak power is 20 W. The signal's average power is 35 mW and peak power is 17.5 W. We use a half waveplate (HWP) and a quarter waveplate (QWP) with polarizing beam splitters (PBS's) on both arms to select the horizontal polarization for the pump and signal beams. The beams are then magnified by telescopes to 2.6 mm FWHM for the signal and 2.8 mm for the pump. Afterwards, they are incident on SLMs (Santec SLM-100) with the angle of incidences of $50^{\circ}$ and $55^{\circ}$, respectively \cite{Santosh19}. The total phase pattern for converting the input Gaussian beam into the $LG_l^p$ beam is given as $\Theta(r,\phi) = -l\phi +\pi \theta (-L_p^{|l|}(\frac{2r^2}{\omega_0^2}))$, with $\theta$ as a unit step function. The phase value is wrapped in the interval between 0 and $2\pi$ to express on the SLMs. 

The signal and pump beams are then combined at a beam splitter (BS) and focused ($f$=200 mm) inside a temperature-stabilized PPLN crystal with a poling period of 19.36 $\mu$m (5 mol.\% MgO doped PPLN, 10 mm length, 3 mm width, and 1 mm height from HC Photonics) for frequency conversion process. HWP is used to ensure the vertically polarized light parallel to the crystal's optical axis. The beam waist of the signal and  pump, both in Gaussian spatial modes, inside the crystal are 45 $\mu$m and 41 $\mu$m, respectively. We also fine tune the time delay between the signal and pump arm to obtain good temporal overlap of their pulses. The output is then passed through two short-pass filters to select the SF light and remove any residual light \cite{QPMS2017}. The central wavelength of the SF light is 775.85 nm with FWHM of 0.05 nm. The other arm of the BS is used to monitor the spatial mode of the pump and signal beams using a near-infrared FIND-R-SCOPE camera. The same arm can also be used to monitor the intensity fluctuation on an high speed oscilloscope (not shown in the figure). The SF light is split in two parts using a flip-able BS. A lens placed at one arm of the flip-able BS is used to image the collimated SF light on a CCD camera with a sensor size of 22.3 mm $\times$ 14.9 mm and a pixel pitch of 4.3 $\mu$m. The other arm of the SF light is coupled into a single mode fiber and detected by the power meter sensor (Thorlabs PM-100D with sensor S130C). This measurement is sent to the computer via a MATLAB interface for the adaptive feedback control process. It updates the phase mask on pump SLM to optimize the selectivity among the signal modes. The effect of the turbulence on the signal is simulated by adding turbulence phase noises to the original phase mask for the LG modes.

\section{RESULTS AND DISCUSSIONS}

\begin{figure}[htbp]
  \centering
   \includegraphics[width=8.5cm]{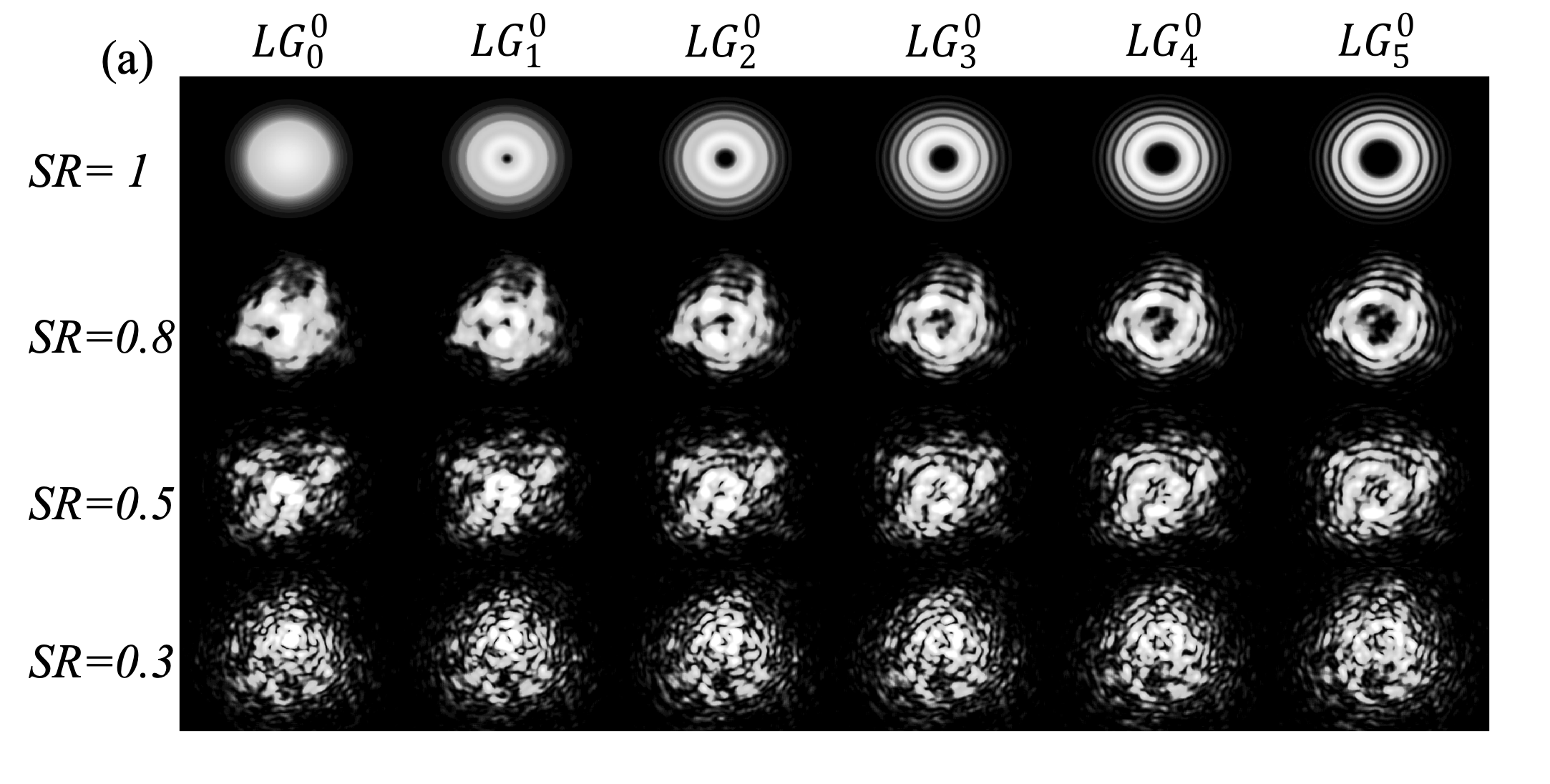}
   \includegraphics[width=8.5cm]{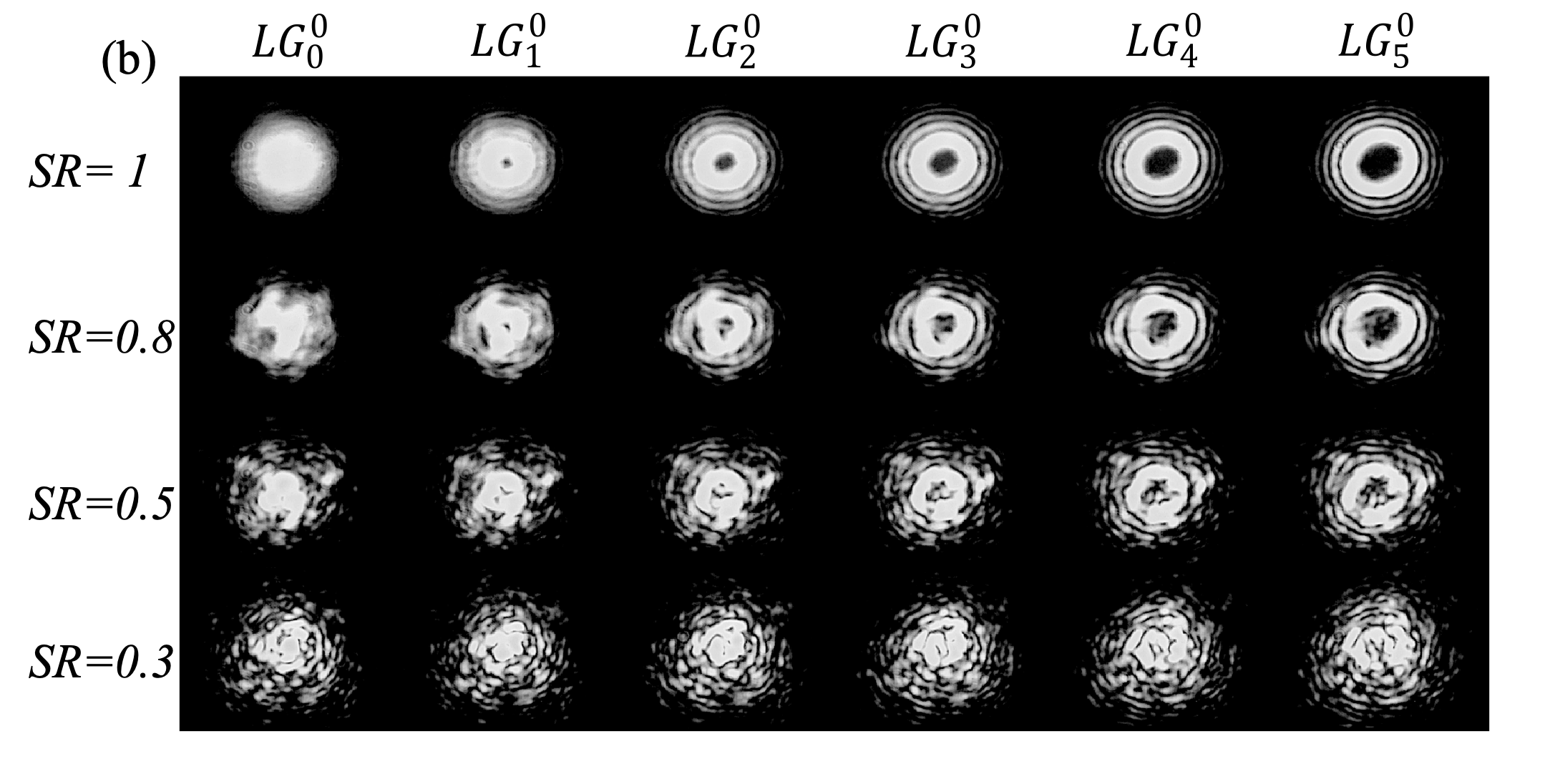}
 \caption{(a) Simulated and (b) measured signal modes without and with varied turbulence.} \label{LGmodes}
 \end{figure}

To illustrate the turbulence effects, in Fig.~\ref{LGmodes} we plot the resulting LG modes at 1558.5 nm under turbulence of different strengths. From Eq.(\ref{SR}), as $SR$ decreases, the strength of the turbulence increases, and the spatial coherence degrades to give distorted intensity distribution. At $SR=0.3$, all modes become illegible. In the figure, the appearance of narrow outer rings is a consequence of the phase only modulation \cite{sephton_revealing_2016}. 

\begin{figure}[htbp]
  \centering
   \includegraphics[width=\linewidth]{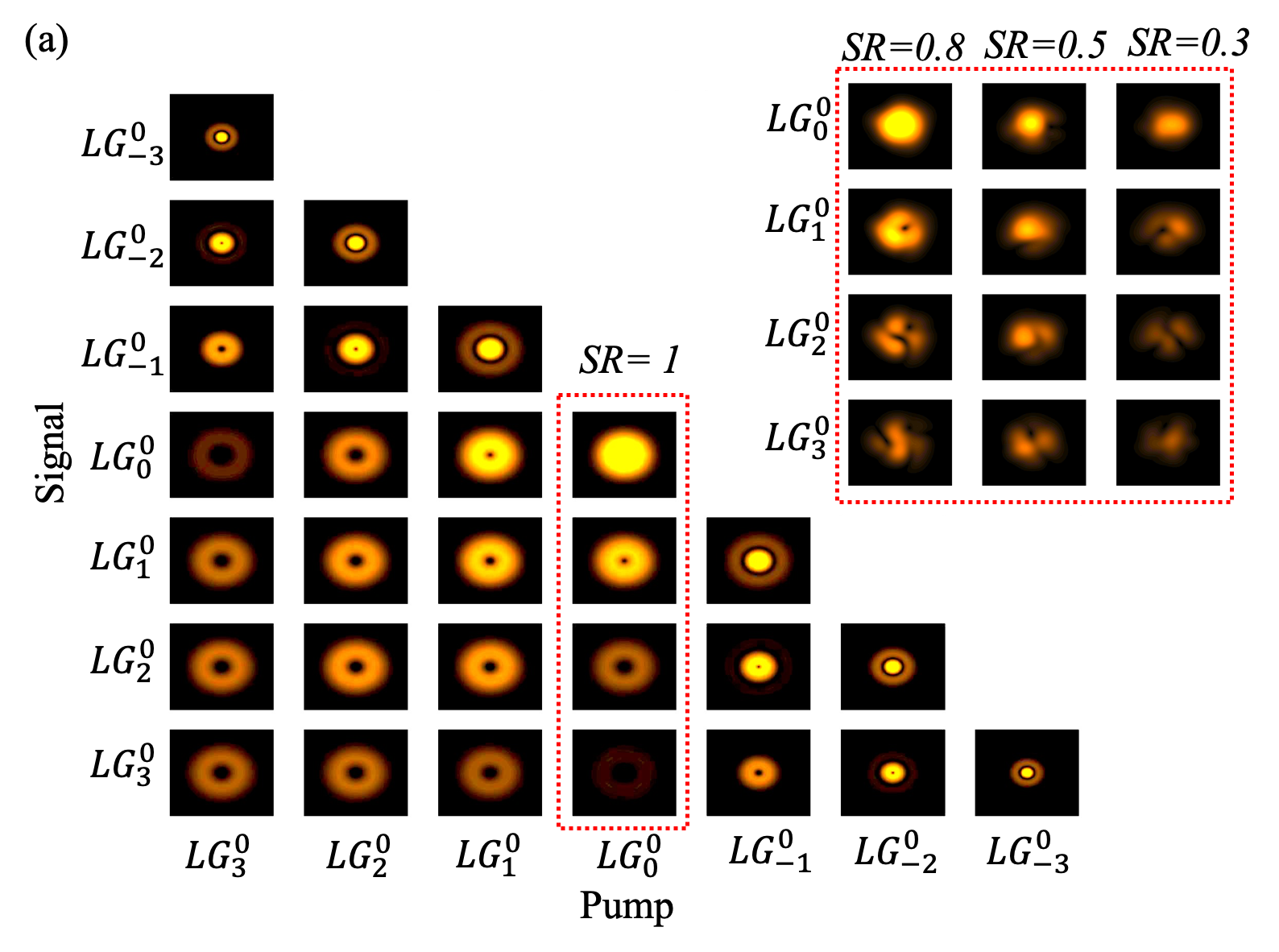}
   
   \includegraphics[width=\linewidth]{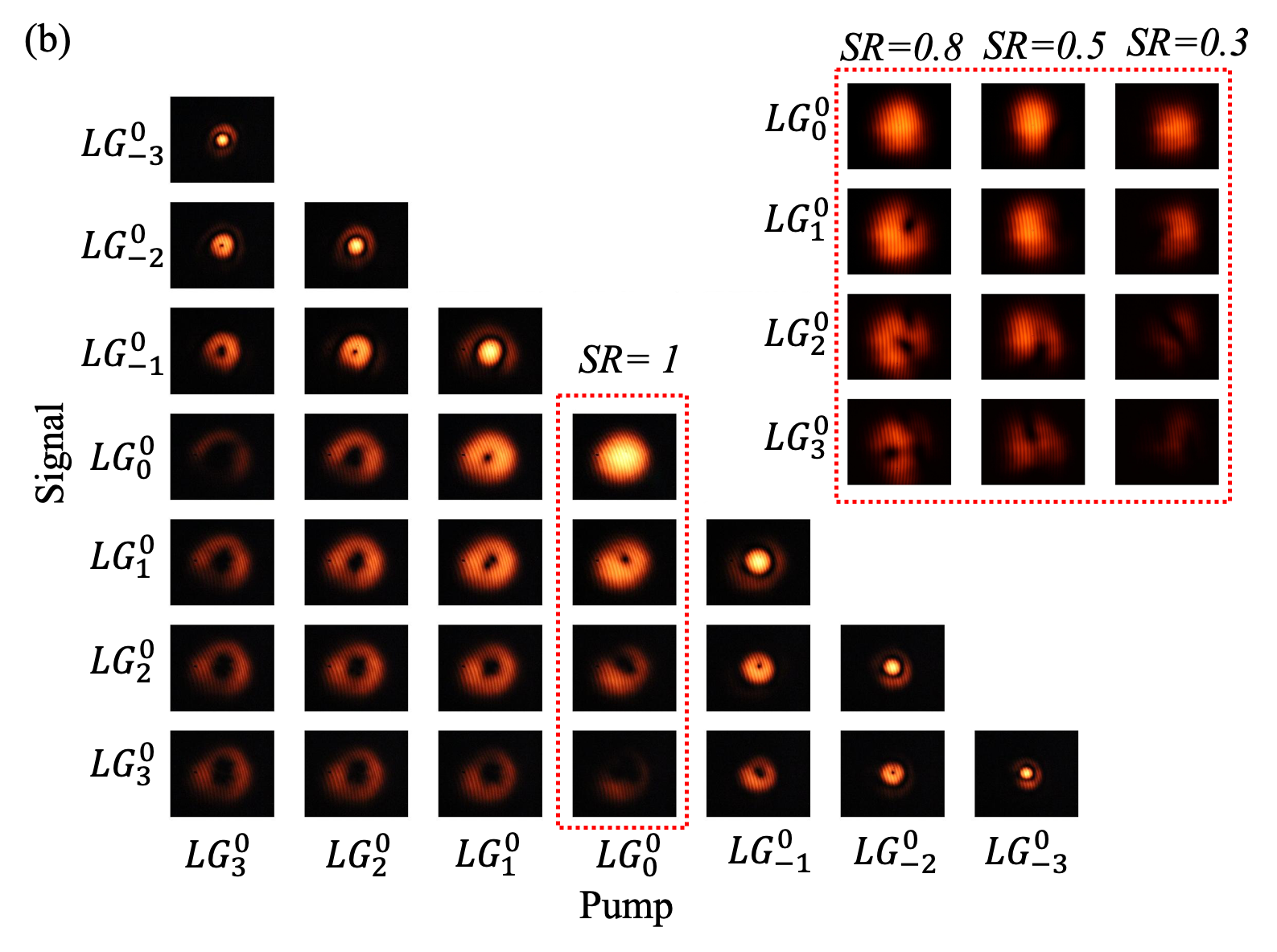}
 \caption{(a) Numerical and (b) experimental results of the generated SF modes with no turbulence on both signal and pump. Insets of (a) and (b) show the SF modes generated by Gaussian pumps but turbulence-distorted signals.} \label{modes}
 \end{figure}

In Fig.~\ref{modes} we show the spatial profiles of the upconverted SF lights for different combinations of signal and pump $LG^p_l$ modes without turbulence ($SR$=1). The numerical results in Fig. \ref{modes} (a) are evaluated by solving Eqs.(\ref{eq2})-(\ref{eq4}) for $|\psi_f(x,y)|^2$. In a previous work, we have studied the selective upconversion of overlapping spatial modes in the non-diffraction regime with the same helicity for the signal and pump beams \cite{Santosh19}. Here, we consider azimuthal indices from $+l$ to $-l$ for both signal and pump beams. When $l_p =-l_s$, the SF mode has a central Gaussian bright spot and thick outer rings, as expected. The Insets of Fig. \ref{modes} show the SF modes for a Gaussian pump with turbulent signal modes. As seen, turbulence significantly reduce the total power and distort the intensity profiles of the generated SF modes. Our simulated SF modes exhibit similar features with the measured results.

Unlike previous studies where the SF power were directly measured at an image plane \cite{Vasilyev14,Santosh19,sephton2019}, here we couple the SF lights into a single mode fiber. The fiber acts as a spatial mode filter on the SF light to improve to detection extinction \cite{gisler_mode-selective_1995,winzer_fiber_1998,wu_mode-selective_2016}, while also providing convenience for subsequent processing of the SF light. For instance, by measuring the total SF power, our previous experiment achieved $\sim$5.6 dB selectivity between two signal modes $LG^0_{0}$ and $LG^0_{1}$ with an un-optimized pump in mode $LG^0_{0}$ \cite{Santosh19}. With the same pump, the extinction is increased to 15.9 dB by measuring the fiber-coupled SF power. In the current setup, the fiber coupling efficiency is $\sim$60\% for the SF light created by a Gaussian signal and a Gaussian pump. In comparison, in a recent study \cite{sephton2019}, the extinction was derived by selecting the only on-axis SF intensity of the central pixel on a CCD camera. Thus the effective detection efficiency is rather low. In contrast, the current setup can achieve both high extinction and high detection efficiency at the same time, which is critical for classical and quantum communications in practical settings. 

To further improve the selectivity, the particle swarm optimization algorithm is used to numerically optimize the pump spatial profiles for the target signal modes \cite{Santosh19}. We then apply these numerically optimized phase masks on SLM to create the pump. However, due to the inevitable phase errors in the SLM and imperfect alignment of the optical beam, the measured selectivity is significantly lower than achievable in simulation. To mitigate the errors and imperfections, we apply an adaptive feedback loop in our experiment to fine tune the pump's phase masks. 

\begin{table}[ht]
 \centering
\arrayrulecolor{gray}%[HTML]{DB5800}

  \centering
\begin{adjustbox}{width=.47\textwidth}
\begin{tabular}{|c|c|c|c|c|c|c|}

\hline 
 \rowcolor[gray]{0.9} {\bf (a)}  & $SR$ & $\xi_1 $ (dB) & $\xi_2$ (dB) & $\xi_3 $ (dB) & $\xi_4$ (dB) & $\xi_5 $ (dB)  \tabularnewline
  \hline 
    \multirow{2}{*}{$S[LG_0^0]^{SR= 1}$}
    & 1  & 27.3 & 25.3 & 32.7 & 35.7 & 44.4  
\\ \cline{2-7}
   & 0.5 & 2.1 & 7.0 & 10.2 & 11.8 & 18.6  
       \tabularnewline
  \hline 

$S[LG_0^0]^{SR=0.5}$&0.5 & 26.1&21.8 &17.8 &23.0 &31.8  
 \tabularnewline

\hline
\end{tabular}%
\end{adjustbox}

\begin{adjustbox}{width=.47\textwidth}
\begin{tabular}{|c|c|c|c|c|c|c|}

\hline 
 \rowcolor[gray]{0.9} {\bf (b)}  &$SR$ & $\xi_1 $ (dB) & $\xi_2 $ (dB) & $\xi_3 $ (dB) & $\xi_4 $ (dB) & $\xi_5 $ (dB)  \tabularnewline
  \hline 

    \multirow{2}{*}{$S[LG_0^0]^{SR=1}$}
    & 1  & 27.2 & 27.6 & 29.6 & 30.5 & 39.1  
\\ \cline{2-7}
    & 0.5 & 4.7 & 9.9 & 14.3 & 15.5 & 19.2  
       \tabularnewline
  \hline 
$S[LG_0^0]^{SR=0.5}$ & 0.5& 20.1&19.2 &18.8 & 20.7&19.3 
 \tabularnewline
 
\hline
\end{tabular}%
\end{adjustbox}

\caption{ (a) Simulated and (b) experimental results for the selective upconversion of the signal modes using optimized pump with ($SR=0.5$) and without turbulence ($SR=1$). $\xi_i$ is the extinction of selecting the $LG^0_0$ over $LG_i^0$ mode. 
} \label{Opt_modes}
\end{table}

Table \ref{Opt_modes} shows the selectivity performance for the target spatial mode $LG_0^0$ over other orthogonal modes, where $\xi_i$ the dB ratio of fiber-coupled SF power of the $LG_0^0$ mode to $LG_i^0$ mode with the same pump. Here, the pump $S[LG_0^0]^{SR}$ is optimized to selectively upconvert signal mode $LG_0^0$ while simultaneously suppressing $LG_1^0$, $LG_2^0$, $LG_3^0$, $LG_4^0$, and $LG_5^0$ without turbulence ($SR$=1), and with the same turbulence noise ($SR$=0.5). In numerical simulations, without turbulence the extinction reaches 44.4 dB, while the best experimental result gives 39.1 dB. As we add the turbulence, the selectivity obtained using the same pump $S[LG_0^0]^{SR=1}$ drops significantly, with the extinction reduced by about 23 dB on average in simulation, and up to 18 dB in experiment. This signifies the strong distortion of the signal modes by the turbulence. To overcome it, we then create re-optimized pump mode $S[LG_0^0]^{SR=0.5}$ using the same feedback method to recover the high extinctions. As a result, the extinctions increase significantly by as high as 24 dB in simulation and 15 dB in experiment. Note that in experiment, while the re-optimization significantly improves the measured extinction for the first LG modes, it gives less improvement on the higher-order LG modes. This is because our current optimization method emphasizes on increasing the lowest extinction to achieve high extinction of the $LG_0^0$ mode over all of the other LG modes. As the extinction against those higher-order modes remain high despite turbulence, the re-optimization improves them only marginally. In practice, higher performance can be achieved by tailoring the optimization to take advantage of any prior information of the modes to be detected. 

\begin{figure}[htbp]
  \centering
   \includegraphics[width=8cm]{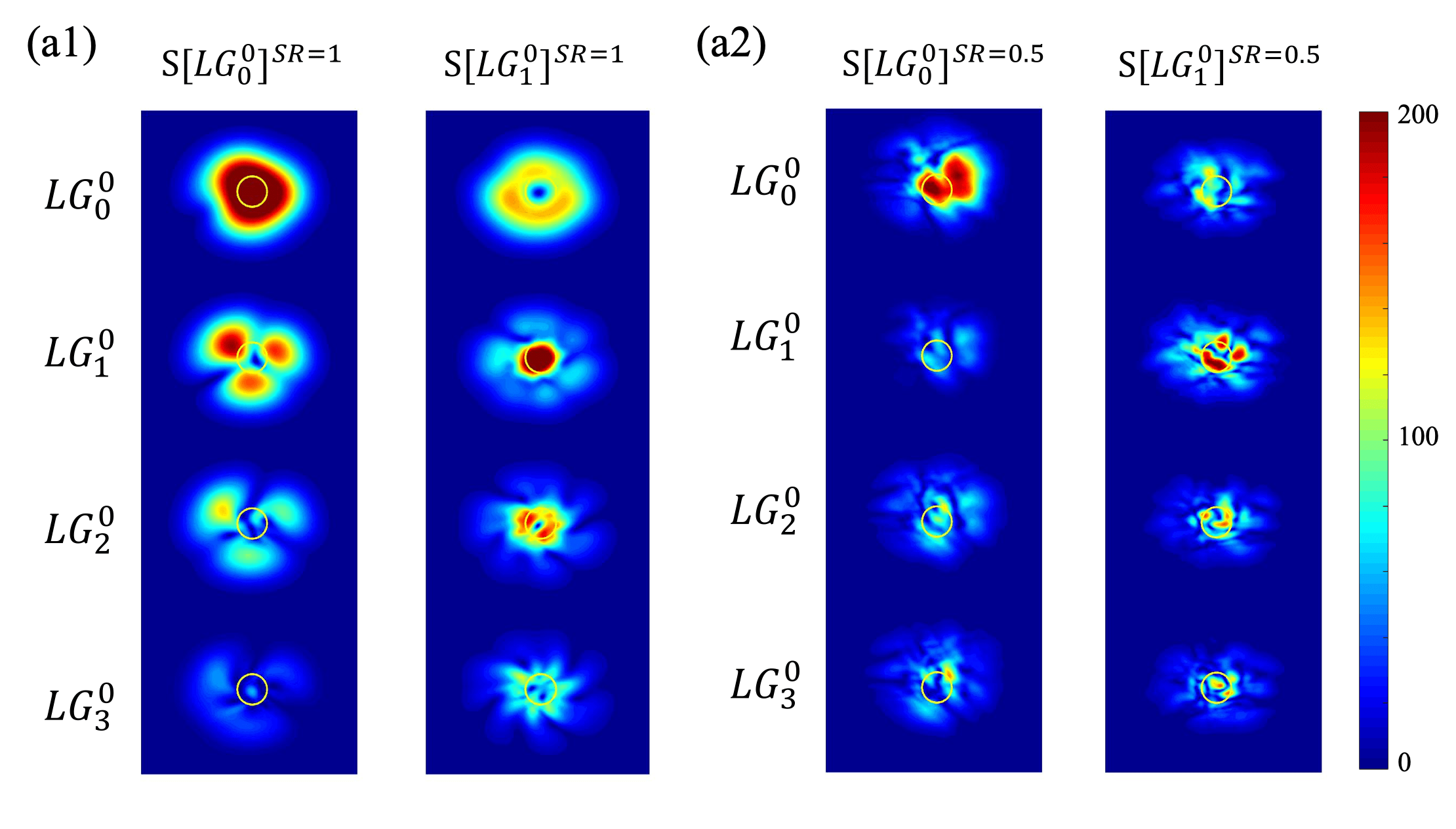}
   \includegraphics[width=8cm]{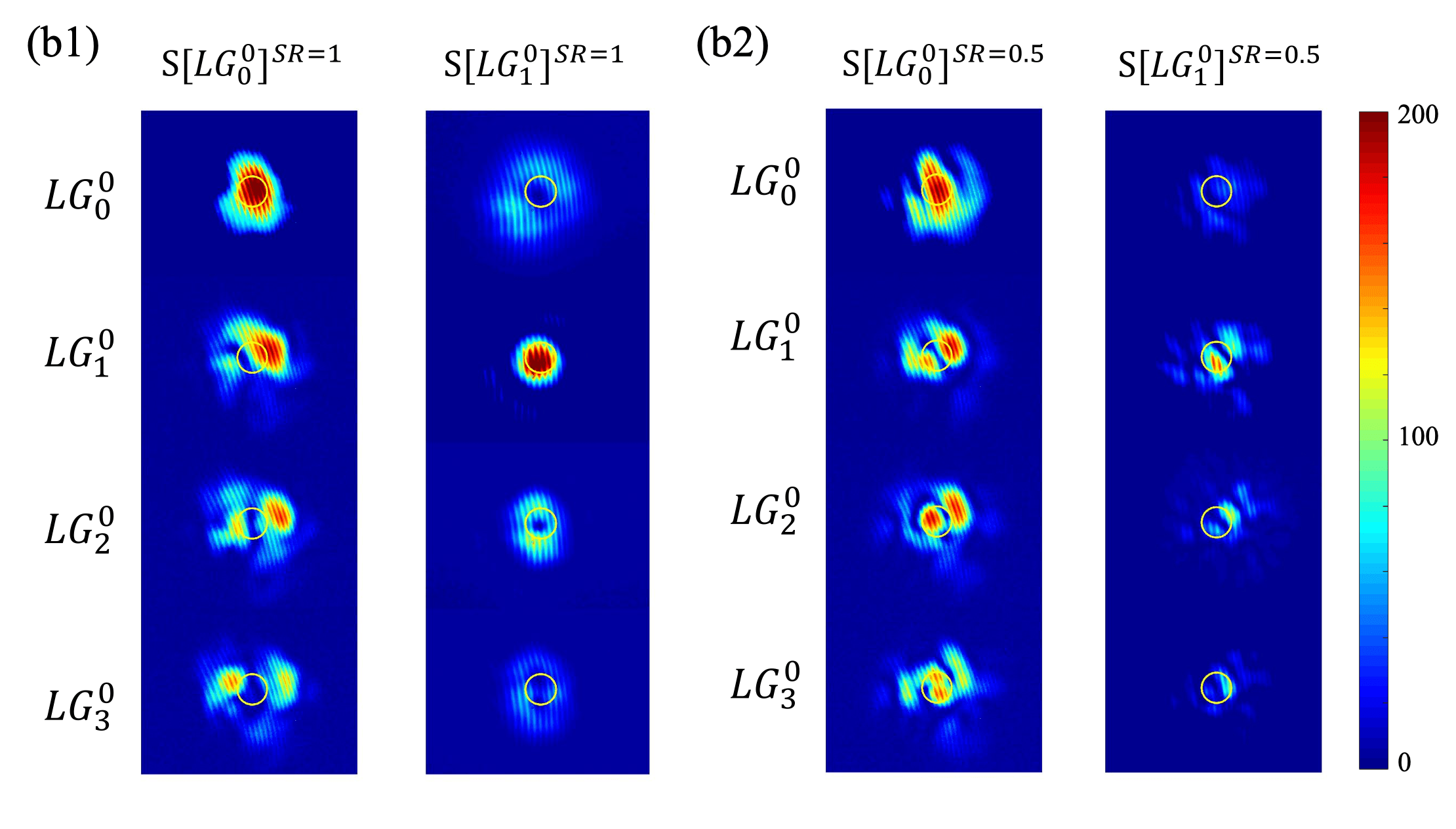}
 \caption{(a1)-(a2) Numerical and (b1)-(b2) experimental results for SF modes created by optimized pumps. In (a1) and (b1), the pump $S[LG_0^0]^{SR=1}$ and $S[LG_1^0]^{SR=1}$ are optimized to selectively upconvert the signal modes $LG^0_0$ and $LG^0_1$, respectively, with no turbulence ($SR=1$). In (a2) and (b2), the pump $S[LG_0^0]^{SR=0.5}$ and $S[LG_1^0]^{SR=0.5}$ are optimized to selectively upconvert signal modes $LG^0_0$ and $LG^0_1$, respectively, with strong turbulence $SR=0.5$. The yellow circle indicated the coupling region into a single mode fiber.} \label{opt-SF}
 \end{figure}

To understand those re-optimization results, in Fig. \ref{opt-SF} we plot the resulting spatial profile of the SF modes with the optimized pump. Here, the optimized pumps $S[LG_i^0]^{SR}$  is designed to selectively upconvert $LG^i_0$ while simultaneously suppressing all others three modes, for the cases of with turbulence ($SR$=0.5) and without turbulence ($SR$=1). The yellow circle highlights the effective region of the SF light coupled into the single mode fiber. As seen, the (re-)optimization significantly increases the optical power inside the fiber-coupled mode area for the target mode, but not the others. Our experimental results for the optimized pump qualitatively agree with the simulations.

\begin{figure*}[b]
  \centering
   \includegraphics[width=12.5cm]{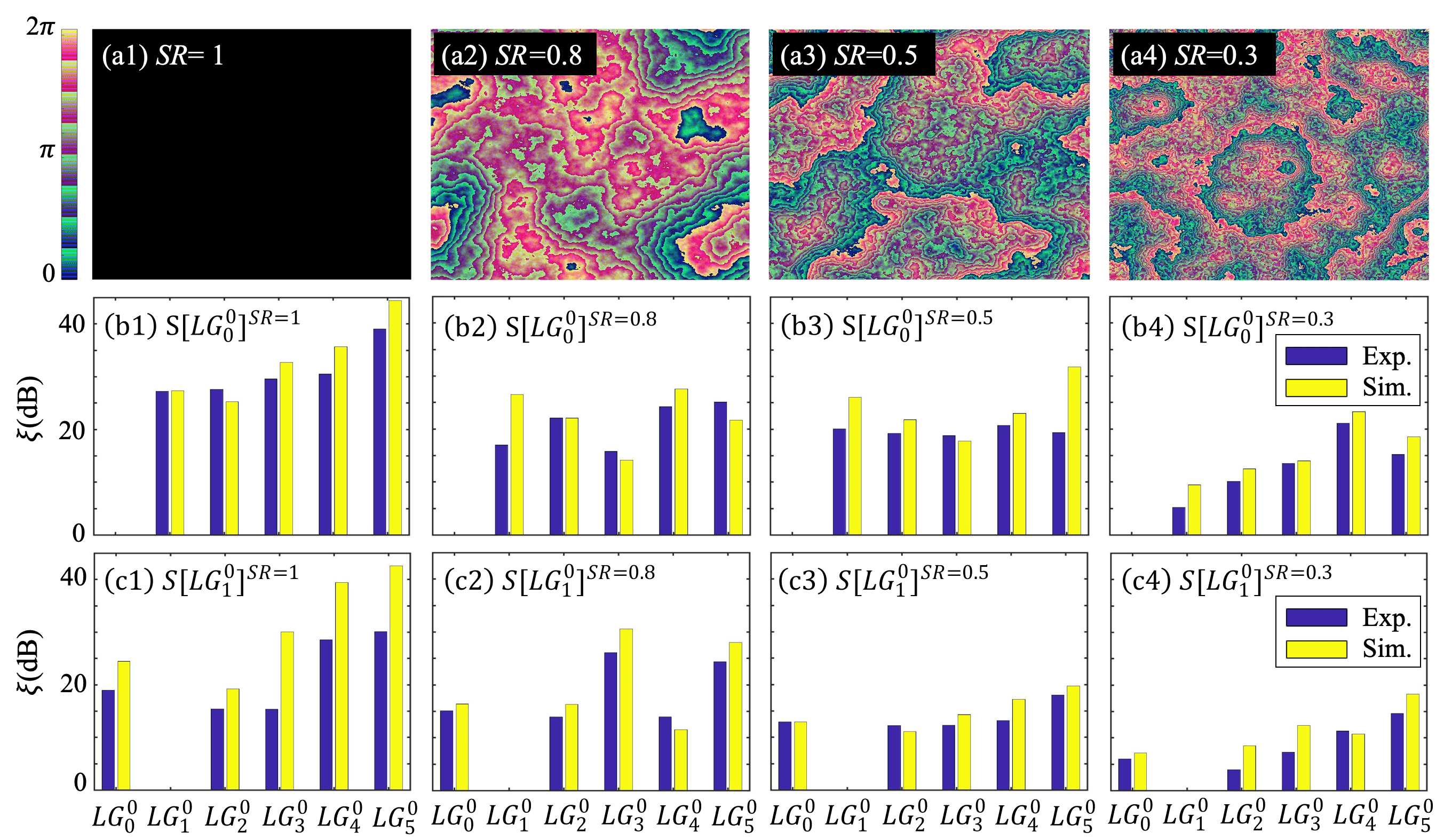}
    \caption{(a1)-(a4)Phase masks to simulate various level of turbulence. (b1)-(b4) Experimental (blue bars) and simulated (yellow bars) results with the (re)-optimized pump $S[LG_0^0]^{SR}$ to select $LG^0_{0}$ signal mode while suppressing other modes under those turbulence. (c1)-(c4) are similar results but to select $LG^0_{1}$. }\label{Optimization}
    \end{figure*}

Figure \ref{Optimization} presents a series of examples on the selective upconversion of spatial modes with re-optimized pump for various turbulence. Figure \ref{Optimization} (a1)-(a4) display the phase masks added to the phase pattern of the LG modes to simulate the turbulence effects. Figure \ref{Optimization} (b) and (c) compare the numerical and experimental extinctions for selectively upconverting $LG_0^0$ and $LG_1^0$ modes, respectively, under various turbulence. As shown in Fig. \ref{Optimization} (b1)-(b4), when the turbulence increases, the extinction drops. However, even for quite strong turbulence at SR=0.3, the average extinction still reaches an average of 16 dB and 13 dB for numeric and experimental results, respectively. For the $LG^1_0$ mode, we obtain similar results, as shown in Fig. \ref{Optimization} (c1)-(c4), with the average extinction drops to 12 dB and 9 dB in simulation and experiment.

\begin{figure*}[htbp]%tbp
  \centering
   \includegraphics[width=12.5cm]{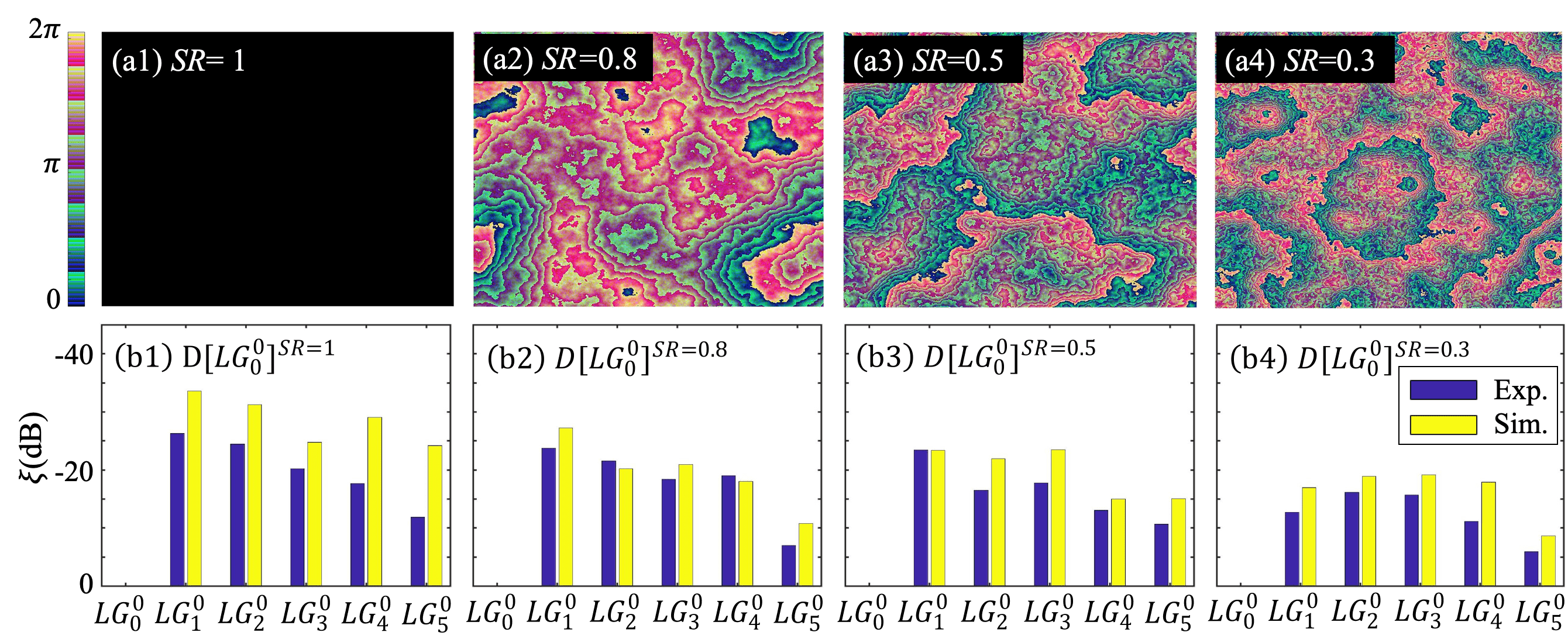}
    \caption{Similar to Fig.\ref{Optimization} but with the pump optimized to de-select the $LG^0_{0}$ signal mode.}\label{Optimization-discard}
\end{figure*}

The above shows the results to selectively upconvert a certain mode over others. With straightforward modification of the convergence condition in the optimization algorithm, the same technique can be employed for other operations, including mode deselection, where a certain mode is dropped, rather than picked, from other overlapping modes. 
Figure \ref{Optimization-discard} shows some de-selection results in the presence of different turbulent strengths. Opposite to $S[LG_0^0]^{SR}$ for selection, now the optimized pump $D[LG_0^0]^{SR}$ is designed to avoid converting one undesired mode while increasing converting other overlapping modes with high efficiency. Figure \ref{Optimization-discard} (b) shows the simulated and measured extinctions for the signal modes by using individually optimized pump modes. Here, a large negative extinction means a much lower conversion efficiency of the target mode versus others, which is desirable. In Fig. \ref{Optimization-discard}, the optimized pump $D[LG_1^0]^{SR=1}$ gives the numerically evaluated extinctions -33.6 dB, -31.2 dB, -24.8 dB, -29.0 dB, and -24.2 dB and experimentally observed extinctions -26.3 dB, -24.4 dB, -20.2 dB, -17.7 dB, and -11.9 dB for the signal modes $LG_1^0$, $LG_2^0$, $LG_3^0$, $LG_4^0$, and $LG_5^0$, respectively. In experiment, the use of re-optimized pump modes for turbulence give extinctions of (b2) -23.7 dB, -21.6 dB, -18.4dB, -19.0 dB, and -7.0 dB for $SR=0.8$; (b3) -23.5 dB, -16.5 dB, -17.7 dB, -13.0 dB, and -10.7 dB for $SR=0.5$; and (b4) -12.7 dB, -16.2 dB, -15.7 dB, -11.1 dB, and -3.9 dB $SR=0.3$, respectively. In comparison, simulated results are (b2) -26.5 dB, -22.1 dB, -14.1 dB, -27.5 dB, and -21.7 dB; (b3) -26.1 dB, -21.8 dB, -17.8 dB, -23.0 dB, and -31.8 dB; (b4) -9.5 dB, -12.5 dB, -14.0 dB, -23.3 dB, and -18.5 dB for $SR =0.8, 0.5$ and $0.3$, respectively. Those results highlight good deselection performance. 
All of our experimental results are in good agreement with the numerical simulations. In the case of strong turbulence, the extinctions are less recovered by pump re-optimization. This is because the initial fiber-coupled SF power is significantly less, so that the feedback control algorithm works less well. This can be improved in the future with more robust algorithms, or by pushing up the conversion efficiency similarly to our previous work in the time-frequency domain \cite{QPMS2017}. These will be subjects of our future efforts.  

\section{Conclusion}
In conclusion, we have numerically and experimentally demonstrated mode selective up-conversion of structured spatial modes mixed with turbulent noise. Unlike any existing adaptive optics method that applies compensating modulation directly to the signal, here the turbulence is accounted for indirectly by modultating the pump modes that drives the upconverison, but not the signal. Hence, it fundamentally eliminates the signal modulation noise or loss, while also allowing exceptional flexibility and capabilities in manipulating and detecting overlapping signals, including quantum light at a single photon level. Our experimental results have achieved $\sim$40 dB extinction under no turbulence, and $\sim$20 dB even in the presence of strong turbulence. The latter is expected to be significantly increased by using better optimization algorithm and higher pump power to enhance the conversion, which will be a subject of our future studies.

We have also demonstrated the selection and de-selection of a particular mode amongst overlapping modes. The same technique applies to superposition modes, as they correspond a new set of modes in different spatial profiles. Favorably, any signal mode that are not converted will remain in its original wavelength and quantum state. They can thus be recycled for cascaded operations to realize complex functionalities. For example, by using a serial or optical-loop setup similar to that in \cite{YuPing13}, efficient quantum state tomography can be implemented on signals in high-order optical angular momentum states \cite{toninelli19}. All of those unique capabilities may prove useful for remote sensing \cite{weimer18}, phase encryption and decryption \cite{Mogensen:00}, compressive imaging \cite{Howland13,Xiaojun11},  quantum free-space optical communication \cite{Tyler09}, and so on.

\section*{Funding}
This research was supported in part by National Science Foundation (Award No. 1842680).

\bibliography{biblio}
\end{document}